\documentclass[showpacs,aps,preprint,graphicx,psfig]{revtex4}
\usepackage{amssymb}
\usepackage{amsmath}
\usepackage{graphicx}
\usepackage{bm}
\def\be{\begin{equation}}
\def\ee{\end{equation}}
\def\bea{\begin{eqnarray}}
\def\eea{\end{eqnarray}}

\begin{document}

\title{Effect of Tensor Correlations on Gamow-Teller States in
$^{90}$Zr and $^{208}$Pb}

\author{C.L. Bai$^{1,2)}$, H. Sagawa$^{3)}$, H.Q. Zhang$^{1,2)}$,
X.Z. Zhang $^{2)}$, G. Col\`o$^{4)}$ and F.R. Xu$^{1)}$}

\affiliation{$^{1)}$School of Physics and State Key Laboratory of Nuclear Physics and Technology, Peking University, China\\
$^{2)}$Department of Nuclear Physics and Beijing Tandem Accelerator National Laboratory,102413 Beijing, China\\
$^{3)}$Center for Mathematical Sciences, University of
Aizu, Aizu-Wakamatsu, Fukushima 965-8560, Japan\\
$^{4)}$Dipartimento di Fisica, Universit$\grave{a}$ degli Studi
and INFN, Sezione di Milano, 20133 Milano, Italy}


\begin{abstract}
The tensor terms of the Skyrme effective interaction are included
in the self-consistent Hartree-Fock plus Random Phase
Approximation (HF+RPA) model. The Gamow-Teller (GT) strength
function of $^{90}$Zr and $^{208}$Pb are calculated with and
without the tensor terms. The main peaks are moved downwards by
about 2 MeV when including the tensor contribution. About $10\%$ of
the non-energy weighted sum rule is shifted to the excitation energy
region above 30 MeV by the RPA tensor correlations.
The contribution of the tensor terms to the energy weighted sum rule is
given analytically, and compared to the outcome of RPA.
\end{abstract}

\maketitle

Nuclei far from the stability line open a new test ground for
nuclear models. Many experimental and theoretical efforts have been
put to study the structure and the reaction mechanisms in the nuclei
near the drip lines. Studies of exotic nuclei far from the
$\beta$-stability line have elucidated unexpected trends for the
shell closures~\cite{Ozawa}. The shell model can explain it if a
finite range tensor force is introduced, which mainly originates
from one-pion exchange~\cite{Otsuka}. In the mean field theory,
almost 30 years ago, the role played by tensor interactions for the
position of the single-particle states was first
discussed~\cite{Stancu} within the HF framework based on Skryme
interactions~\cite{Skyrme}. Then, tensor force was included in the
Skyrme-Landau parametrization and RPA (Random Phase Approximation)
calculation in Ref.~\cite{Liu}. However, the tensor force was
essentially dropped in most Skyrme parameter sets which have been
fitted and which are still widely used in nuclear structure
calculations. Recently, in Ref.~\cite{Brown}, a Skyrme interaction
which includes the tensor contribution was fitted. Then, tensor
terms were added perturbatively in Refs.~\cite{Brink}
and~\cite{Colo} to the existing standard parametrizations
SIII~\cite{Beiner} and SLy5~\cite{Chabanat}, respectively.
Eventually, several new parameter sets have been fitted in
Ref.~\cite{Lesinski} and used for systematic investigations within
the Hartree-Fock-Bogoliubov (HFB) framework. The inclusion of tensor
terms in the Skyrme HF calculations achieved considerable success in
explaining some features of the evolution of single-particle
states~\cite{Wei}. However, there has been, so far, no RPA or QRPA
(Quasiparticle Random Phase Approximation) program available to
study the effect of the tensor terms on the excited states of
nuclei.

The present work is devoted to including the tensor terms of the
Skyrme effective interaction in the self-consistent HF plus RPA
calculations. In particular, we are interested in the GT
transitions, which should be affected in keeping with the fact
that the corresponding operator is
spin-dependent~\cite{Shimizu,Arima}. In the study of GT
transitions, the quenching problem is of some relevance. The
experimentally observed strength from 10 to 20 MeV excitation
energy (with respect to the ground state of the target nuclei) is
about $50\%$ of the model-independent non-energy weighted sum rule
(NEWSR); this percentage becomes about $70\%$ if the whole
strength in the neighboring energy region is
collected~\cite{Rapaport}. We deem interesting to study if the
tensor force has an effect in shifting the strength already at one
particle-one hole (1p-1h) level. Coupling the GT with two
particle-two hole states is essential to describe the resonance
width but it is not expected to affect strongly the position of
the main GT peak; the effect of the tensor force in connection
with the 2p-2h coupling was studied in Ref~\cite{Bertsch}.

In this letter, we employ the triplet-even and triplet-odd
zero-range tensor terms, which have the form originally postulated
in the pioneering work by Skyrme and read~\cite{Stancu,Skyrme}
\begin{equation}
\begin{array}{llll}
\displaystyle\\V^T&=&\displaystyle\frac{T}{2}\{[(\bf\sigma_1\cdot\bf{k}^{\prime})
(\bf\sigma_2\cdot\bf{k}^{\prime})-\frac{1}{3}\left(\sigma_1\cdot\sigma_2\right)\bf{k}^{\prime2}]\delta\left({\bf r}_1-{\bf r}_2\right)\\
&&+\delta({\bf r}_1-{\bf r}_2)\left[(\bf\sigma_1\cdot\bf{k})(\bf\sigma_2\cdot\bf{k})
-\frac{1}{3}\left(\sigma_1\cdot\sigma_2\right)\bf{k}^{2}\right]\}\\
&&+\displaystyle\frac{U}{2}\{\left(\sigma_1\cdot{\bf k}^{\prime}\right)\delta\left({\bf r}_1-{\bf r}_2\right)(\sigma_2\cdot\bf k)
+\left(\sigma_2\cdot\bf{k}^{\prime}\right)\delta({\bf r}_1-{\bf r}_2)(\sigma_1\cdot\bf{k})\\
&&-\frac{2}{3}\left[({\bf \sigma}_1\cdot{\bf \sigma}_2){\bf k}^\prime\cdot\delta({\bf r}_1-{\bf r}_2)\bf{k}\right]\}
\end{array}
\end{equation}
In the above expression, the operator ${\bf k}=\left(\bf
\nabla_1-\bf \nabla_2\right)/2i$ acts on the right and ${\bf
k}^\prime=-\left(\bf \nabla_1^{\prime}-\bf
\nabla_2^{\prime}\right)/2i$ acts on the left. The coupling
constants T and U denote the strengths of the triplet-even and
triplet-odd tensor interactions, respectively. The calculation
employs, consistently with the choice of the Skyrme force SIII and
with Ref.~\cite{Brink}, the values $T=1008$ MeV fm$^5$ and
$U=-432$ MeV fm$^5$. Similar values of the tensor interactions
have been suggested in Ref.~\cite{Colo} in the study of
spin$-$splitting of Sb isotopes. In Refs.~\cite{Brink,Colo}, the
parameters T and U are chosen in such a way to improve the
absolute values and the isotopic(or isotonic) trends of
single-particle energies. This criterion limits the possible
choice on both the magnitude and sign of T and U. In this way, one
obtains an effective tensor interaction which does not necessarily
correspond to the result of a G-matrix calculation, since effects
from three-body force and nuclear correlations can have been
effectively included. This point has been discussed in
Ref.~\cite{Brown}.

The main effect of the tensor terms on HF calculations is a
modification of the spin-orbit potential(the total binding energies
and radii being, as a rule, less affected). The spin-orbit potential
is given by
\begin{eqnarray}
U_{S.O.}^{(q)}=\frac{W_0}{2r}(2\frac{d\rho_q}{dr}+\frac{d\rho_{q\prime}}{dr})
+(\alpha\frac{J_q}{r}+\beta\frac{J_{q\prime}}{r}),
\end{eqnarray}
In this expression, q=0(1) labels neutrons (protons).
$J_n$ and $J_p$ are the so-called spin-orbit densities
of neutrons and protons
respectively. Their definition can be found in
Ref.~\cite{Vautherin}.
The first
term in the r.h.s comes from the Skyrme two-body spin-orbit
interaction, whereas
the second term includes both a central exchange and a tensor
contribution, that is, $\alpha=\alpha_C+\alpha_T$ and
$\beta=\beta_C+\beta_T$ with
\begin{eqnarray}
\alpha_C&=&\frac{1}{8}(t_1-t_2)-\frac{1}{8}(t_1x_1-t_2x_2),\\
\beta_C&=&-\frac{1}{8}(t_1x_1+t_2x_2),\\
\alpha_T&=&\frac{5}{12}U,\\
\beta_T&=&\frac{5}{24}(T+U)
\end{eqnarray}
It should be noted that $J_q$ gives essentially no contribution
in the spin-saturated cases.
Therefore, we choose $^{90}$Zr and $^{208}$Pb as examples to be
calculated. $^{90}$Zr is a proton spin-saturated nucleus, with a
spin-unsaturated neutron orbit $1g_{9/2}$. $^{208}$Pb is chosen as
it is not saturated either in protons or neutrons. The two examples
should allow elucidating separately the role of triplet-even and
triplet-odd terms.

The HF plus RPA model is described in many textbooks and papers;
accordingly, we give only few details about our numerical
implementation. We start by solving the HF equations in coordinate
space with a radial mesh extending up to 20 fm in a step of 0.1 fm.
When the Skyrme HF potential is calculated, the single-particle
energies and wave functions of the occupied and unoccupied levels
can be solved by using an expansion over a harmonic oscillator
basis. This basis is large enough to ensure that our results are
stable and it extends up to a maximum value of the main quantum
number $N_{max}=$10 and 12, for $^{90}$Zr and $^{208}$Pb,
respectively.

Since the tensor force is spin-dependent and affects the spin-orbit
splitting, the spin mode is very likely to receive strong influence.
we study hereafter the GT excitation as the well-known spin mode.
The operator for GT transitions is defined as
\begin{eqnarray}
\hat{O}_{GT\pm}&=&\sum\limits_{im}t^i_{\pm}\sigma_m^i
\end{eqnarray}
in terms of the standard isospin operators,
$t_\pm=\frac{1}{2}(t_x{\pm}it_y)$. In the charge-exchange RPA, the
$t_-$ and $t_+$ channels are coupled and the corresponding
eigenstates emerge from a single diagonalization of the RPA
matrix.

In self-consistent charge-exchange HF+RPA calculations, the NEWSRs
$m_\pm(0)$ and the Energy-Weighted Sum Rules (EWSR) $m_\pm(1)$
(associated with the two different isospin channels) satisfy the
following relations
\begin{equation}
\begin{array}{lll}
m_-(0)-m_+(0)&=&\sum\limits_\nu\left( \vert\langle\nu\vert
O_-\vert 0\rangle\vert^2-\vert\langle\nu\vert
O_+\vert 0\rangle\vert^2 \right)\\
&=& \langle 0\vert \left[ O_-, O_+ \right] \vert 0\rangle,
\end{array}
\label{diffNEWSR}
\end{equation}
\begin{equation}
\begin{array}{lll}
m_-(1)+m_+(1)&=&\sum\limits_\nu \left( \vert\langle\nu\vert
O_-\vert 0\rangle\vert+\vert\langle\nu\vert
O_+\vert 0\rangle\vert^2 \right) E_\nu\\
&=& \langle 0\vert \left[ O_+, \left[ H, O_- \right] \right]
\vert 0\rangle,
\end{array}
\label{sumEWSR}
\end{equation}
where $O_+$ ($O_-$) is a generic charge-changing operator
proportional to $t_+$ ($t_-$).
In the GT case, the difference of NEWSRs (\ref{diffNEWSR})
is model-independent and turns out to be
\begin{eqnarray}
m_-(0)-m_+(0)=3(N-Z),
\end{eqnarray}
The sum of the EWSRs(\ref{sumEWSR}) is model-dependent and it
receives a contribution from the tensor interaction, which is
obtained by replacing the total Hamiltonian $H$ in the double
commutator of (\ref{sumEWSR}) with $V^T$. If there is enough
neutron excess, and the contributions from the $t_+$ channel to
the sum rules, $m_+(0)$ and $m_-(0)$, are small, then we can
estimate the effect of the tensor interaction on the GT centroid
in the $t_-$ channel by writing
\begin{equation}
\begin{array}{lll}
\Delta E_{GT}&=&\frac{m_-(1)}{m_-(0)}\\
&\sim&\frac{m_-(1)+m_+(1)}{m_-(0)-m_+(0)}\\
&=&\frac{4\pi}{3(N-Z)}\int
drr^2[-(\frac{5}{2}U+\frac{5}{6}T)J_nJ_p-\frac{5}{3}U(J_n^2+J_p^2)],
\end{array}
\label{shift}
\end{equation}
where the last line comes from a lengthy but straightforward
evaluation of the double commutator.

In the present work, we do not include the two-body spin-orbit
residual interaction in RPA. Consequently our calculations are
not, strictly speaking, fully self-consistent. However, this
term of the residual interaction has been shown to be very
small~\cite{Fracasso} in the case of the GT. Therefore,
we can claim that self-consistency is not seriously broken.
We do not make any further approximation, and, in particular,
we include in HF the central exchange terms associated with
$\alpha_C$ and $\beta_C$.

Only the values reported in Table~\ref{Table1} are, however,
calculated by dropping completely the spin-orbit contribution, both
at HF and RPA level. This calculation (with the Skyrme parameter
$W_0$ set at 0) is not expected to be compared with the experimental
findings but respects self-consistency in a strict sense. The shift
in the GT centroid caused by the inclusion of tensor terms,
[calculated by using either RPA or the analytical formula
(\ref{shift})], and the EWSR $m_-(1)+m_+(1)$ obtained from RPA, are
listed in Table~\ref{Table1} for the two nuclei $^{90}$Zr and
$^{208}$Pb. One should notice the good agreement between the RPA
results and the analytical results for the shift.

\begin{figure}[tbp]
\includegraphics[scale=1.4,clip]{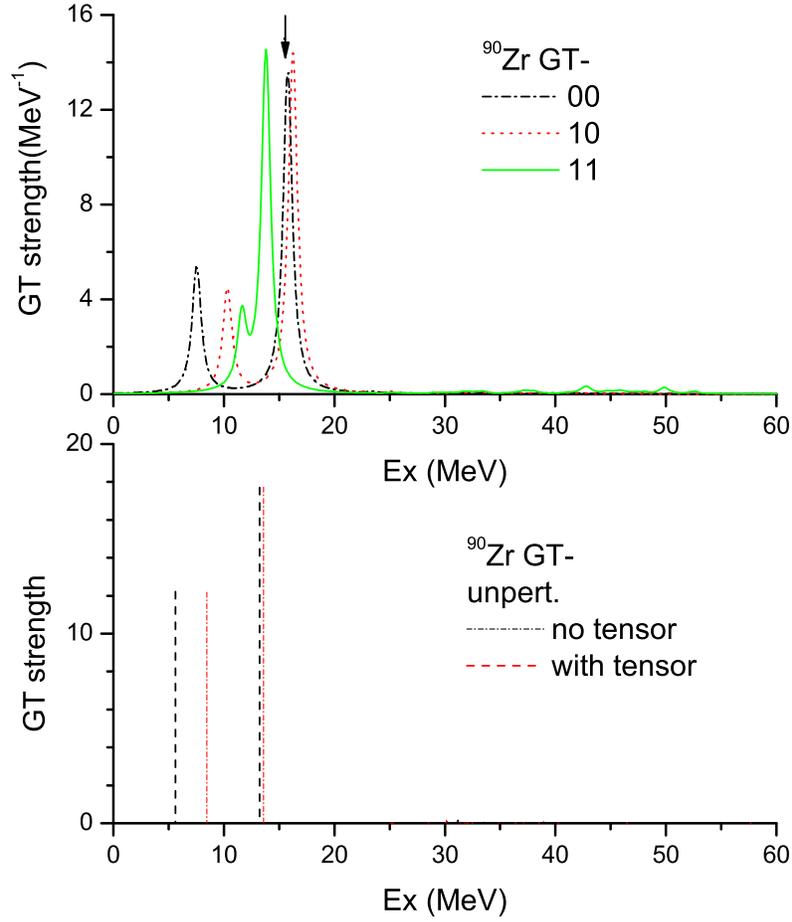} 
\caption{The GT$_-$ strength in $^{90}$Zr. In panel (a), the RPA
results are displayed, by smoothing them with Lorentzians having 1
MeV width. As explained in the text, result labelled by 00
corresponds to neglecting the tensor terms in both HF and RPA; 10
corresponds to including the tensor terms in HF but neglecting them
in RPA; finally, 11 corresponds to including the tensor terms in
both HF and RPA. The arrow denotes the experimental energy. In panel
(b), the unperturbed strength is shown. } \label{fig1}
\end{figure}
\begin{figure}[tbp]
\includegraphics[scale=1.4,clip]{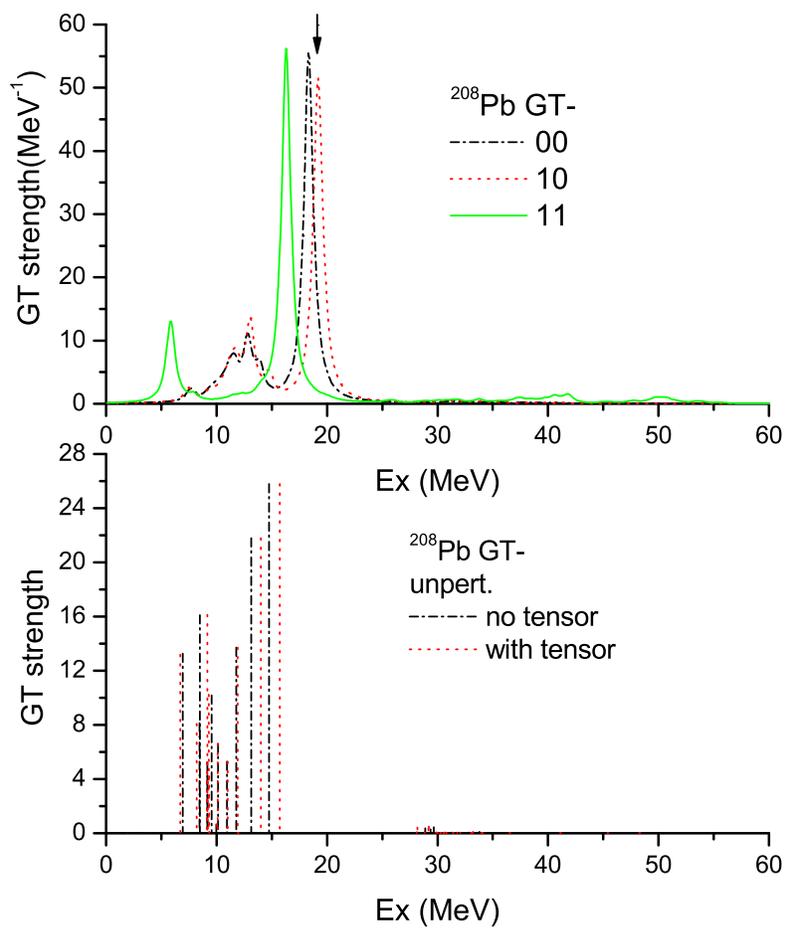} 
\caption{The same as Fig.~\ref{fig1} in the case of $^{208}$Pb.
}
\label{fig2}
\end{figure}

The GT$_-$ strength distributions in $^{90}$Zr and $^{208}$Pb are
shown in Figs.~\ref{fig1} and~\ref{fig2}, respectively. The
calculated results are smoothed by averaging the sharp RPA peaks
with Lorentzians weighting function having 1 MeV width. The tensor
force affects these results in two ways. Firstly,  it changes the
single-particle energies (s.p.e.) in the HF calculation; secondly,
it contributes to the RPA residual force. We do three different kind
of calculations to analyze separately these effects. In the first
one, the tensor terms are not included at all. In the second one, we
include tensor terms in HF but drop them in RPA. This calculation is
not self-consistent, but it displays the effects of changes in
single-particle energies on the strength distribution. In the last
one, the tensor terms are included both in HF and RPA calculations.
For simplicity, results of the three categories of calculations are
labeled by 00, 10 and 11, respectively.

We have evaluated the amounts of NEWSR $m_-(0)$ and EWSR $m_-(1)$
in different excitation energy regions, and listed them in
Table~\ref{Table2}. When the tensor terms are not included in the
residual interaction (i.e., the calculations labeled by 00 and
10), the values of NEWSR in the energy region between 30-60 MeV
for both $^{90}$Zr and $^{208}$Pb are small, namely few percent of
the total NEWSR (see the Figs. 1(a) and 2(a)). But in the case 11,
about $10\%$ of NEWSR is shifted from the lower energy region to
the higher energy region. Moreover, we can see that essentially no
unperturbed strength appears in this region (see the Figs. 1(b)
and 2(b)). This means that including tensor terms in simple RPA
calculation shifts about $10\%$ of the GT strength to the energy
region 30-60 MeV. While 2p-2h couplings will increase further
these high energy strength, we would like to stress that the
tensor correlations move substantial GT strength from the low
energy region 0-30 MeV to the high energy region 30-60 MeV even
within the 1p-1h model space.

The EWSR in the energy region below 30 MeV is of course also
decreased after the inclusion of the tensor terms. From
Table~\ref{Table2}, we see that an appreciable amount of EWSR
(that is, 25\% and 29\% of EWSR in $^{90}$Zr and $^{208}$Pb,
respectively) is shifted from the lower energy region (0-30 MeV)
to the higher energy region (30-60 MeV) by including tensor terms
in HF plus RPA calculations.

In $^{90}$Zr, from Fig 1(a) one can notice that the GT strength is
concentrated in two peaks in the region below 30 MeV. There are
only two important configurations involved which are
$(\pi1g_{9/2}-\nu1g^{-1}_{9/2})$ and
$(\pi1g_{7/2}-\nu1g^{-1}_{9/2})$ (see Fig 1(b)). When the tensor
terms are included only in HF and neglected in RPA, the centroid
in the energy region of 0-30 MeV is moved upwards by about 1.5
MeV, and the high energy peak at E $\sim16$ MeV is moved upwards
by only 0.5 MeV, as compared with the results without tensor
terms. When the tensor terms are included in both HF and RPA, the
centroid of the GT strength in the energy region 0-30 MeV is moved
downwards by about 1 MeV, and the high energy peak is moved
downwards about 2 MeV, as compared with the results obtained
without tensor terms. Including tensor terms in RPA makes the two
main separated peaks closer (this situation also happens for
$^{48}$Ca). This result can be attributed from the HF and RPA
correlations of the tensor terms. From the typical effect of the
tensor correlations on HF field~\cite{Otsuka,Colo}, when the
$\nu1g_{9/2}$ orbit is filled by neutrons, the tensor
 correlations give a quenching on the spin$-$orbit splitting between
 $\pi1g_{9/2}$ and  $\pi1g_{7/2}$ orbits
  so that the unperturbed energies of the two
main $p-h$ configurations ~$(\pi1g_{7/2}-\nu1g^{-1}_{9/2})$ and
 $(\pi1g_{9/2}-\nu1g^{-1}_{9/2})$ are closer in energy as it is shown
  in Fig. 1(b).
The RPA results in Fig. 1(a) labelled by (00) and (10) reflect
these changes of HF single particle energies due to the tensor
correlations and the energy difference between two peaks is
narrower. Meanwhile, the RPA correlation associated with tensor
terms move the higher energy peak downwards, and this effect can
be seen in the results in Fig.1(a) labelled by (10) and (11). For
GT transitions in the energy region of 30-60 MeV, several dominant
configurations are expected and they receive some strength from
the low excitation energy region due to tensor correlations.

In $^{208}$Pb, from Fig 2(a) we see that the GT strength is
concentrated in two peaks in the low energy region of 0-30 MeV.
There are eleven important configurations which do contribute to
these peaks. When the tensor terms are only included in HF and
neglected in RPA, the centroid of these peaks is moved upwards
about 0.5 MeV, and the higher energy peak at E $\sim$ 18 MeV is
also raised by about 0.8 MeV. When  the tensor terms are included
in both HF and RPA calculation, the centroid of these peak  moves
downwards by about 1.5 MeV, and the higher energy peak moves also
downwards by about 3.3 MeV, compared with the result obtained
without tensor terms. By including tensor terms in the RPA
calculation, the GT strengths in the energy region of 30-60 MeV
are increased substantially by the shift of the strength in the
energy region of 0-30 MeV through the tensor force.

We have calculated the GT strength in $^{90}$Zr by adding the
presently used tensor terms to SGII and obtained the same result
that about $10\%$ of the NEWSR appears in the high energy region
of 30-60 MeV.

In conclusion, we have studied the effect of the tensor
correlations on the GT excitations in $^{90}$Zr and $^{208}$Pb in
the HF plus RPA framework with the Skyrme interaction SIII. If the
tensor terms are included only in HF but neglected in RPA, the
strength distribution is only slightly shifted to higher energy.
But if the tensor terms are included in both HF and RPA,
  the centroid of GT strength  in
the energy region below 30 MeV is moved downwards by about 1 MeV
for $^{90}$Zr and 3.3 MeV for $^{208}$Pb.  At the same time,
 the dominant  peak at  E $\sim$ 16 MeV (18 MeV)
in $^{90}$Zr ($^{208}$Pb)
 is also moved downwards by about 2 MeV (3 MeV).
It is pointed out for the first time that about $10\%$ of NEWSR is
moved to the high energy region of 30-60 MeV by the tensor
correlations in RPA even within the 1p-1h model space. We give the
analytical formula to estimate the effect of the tensor force on
the mean GT energy. These formulas predict the upwards energy
shift of the average excitation energy due to the tensor
correlations.  It agrees quite well with our numerical RPA
results. It is interesting to point out that  the main GT peak,
contrarily, gets an energy shift downwards because of the peculiar
features of the tensor correlations. In fact, the upwards shift of
the average energy
 is the outcome of the
GT strength appearing in the high energy region between 30-60 MeV,
 but does not correspond to the energy shift of main GT peak.
 Since the tensor interaction is spin-dependent,
we expect that it can have important effects not only on the GT
transitions, but on spin-dipole and other spin dependent
excitation modes as well. These issues will be discussed in a
forthcoming paper.
\begin{acknowledgments}
We thank for useful discussion with Prof. Z.Y.Ma. This work is
supported by the Natinal Natural Science Foundation of China under
Grant Nos 10275092 and 10675169 and the National Key Basic Research
Program of China under Grant No.2007CB815000 and also in part by the
Japanese Ministry of Education, Culture ,Sports, Science  and
Technology
  by Grant-in-Aid
for Scientific Research under
 the program number (C (2)) 20540277 ..

\end{acknowledgments}

\clearpage

\begin{table}[hbt] \centering
\caption{Values of the EWSR $m_-(1)+m_+(1)$ obtained from
self-consistent HF plus RPA calculations with and without the tensor
terms. $\delta E_{RPA}$ and $\delta E_{DC}$ are the contributions of
the tensor terms to the GT centroid energy calculated, respectively,
by using RPA and the analytical formula(\ref{shift}). In the case of
the numbers reported here (not for the other results in this paper),
the spin-orbit term is dropped both at HF and RPA level. See also
the main text. \label{Table1}}
\begin{tabular}{ccccc}
\hline\hline
   & $m_-+m_+(1;{\rm no\ tensor})$ & $m_-+m_+(1;{\rm with\ tensor})$
   & $\delta E_{RPA}$ & $\delta E_{DC}$\\
   & MeV & MeV & MeV & MeV \\
\hline \hline
 $^{90}Zr$ & 271.45 & 338.68 & 2.241 & 2.276 \\
 $^{208}Pb$ & 1854.12 & 2000.76 & 1.111 & 1.118\\
\hline
\end{tabular} \thinspace
\end{table}
\clearpage

\begin{table}[hbt] \centering
\caption{Values of the NEWSR $m_-(0)$ and EWSRs $m_-(1)$ for
$^{90}$Zr and $^{208}$Pb in different excitation energy regions. The
two-body spin-orbit interaction is included in HF but neglected in
RPA calculation. The results labeled by 00 correspond to neglecting
the tensor terms both in HF and RPA; 10 corresponds to including the
tensor terms in HF but neglecting them in RPA; 11 corresponds to
including the tensor terms both in HF and RPA. See the text for a
discussion of the effects of the tensor terms. \label{Table2}}
\begin{tabular}{c|ccccccc}
\hline\hline
& type of     & $m_-(0)$ & $m_-(0)$ & $m_-(1)$ & $m_-(1)$  & $m_-(1)$ & $m_+(1)$ \\
& calculation & 0-30MeV  & 30-60MeV  & 0-30 MeV & 30-60 MeV & total    & total \\
\hline \hline
              & 00 & 29.16 & 0.71 & 395 & 26.2 & 421.8 & 10.1\\
$^{90}$Zr     & 10 & 29.16 & 0.79 & 444 & 22 & 466 & 11.1\\
              & 11 & 27.00 & 2.89 & 366.9 & 122 & 493.2 & 10.3\\
\hline
              & 00 & 127.54 & 3.43 & 2080 & 124.5 & 2212.8 & 18.8\\
$^{208}$Pb    & 10 & 127.38 & 3.68 & 2176 & 93 & 2269 & 21\\
              & 11 & 114.10 & 16.58 &1658 & 694 & 2370 & 19.3\\
\hline
\end{tabular}\thinspace
\end{table}


\end{document}